\documentclass[prb,reprint,showpacs,showkeys]{revtex4-1}
\usepackage{graphicx}
\usepackage{amsmath,amssymb}
\usepackage{bm}
\usepackage{hyperref}
\newcommand{\eb}{\varepsilon_\mathrm{B}}
\newcommand{\ee}{\varepsilon_0}
\newcommand{\cc}{\mathrm{c.c.}}

\newcommand{\sech}{\mathrm{sech}}
\newcommand{\im}{\mathrm{Im~}}
\newcommand{\re}{\mathrm{Re~}}
\newcommand{\rmx}{\mathrm{x}}
\newcommand{\rmr}{\mathrm{r}}
\newcommand{\rmb}{\mathrm{b}}
\newcommand{\rms}{\mathrm{s}}

\begin{document}
\title{Resonant modulational instability and self-induced transmission effects in semiconductors: Maxwell-Bloch formalism}
\author{Oleksii A. Smyrnov}
\email{oleksii.smyrnov@mpl.mpg.de}
\author{Fabio Biancalana}
\affiliation{Nonlinear Photonic Nanostructures Group, Max Planck Institute for the Science of Light, G\"{u}nther-Scharowsky
Stra{\ss}e 1/26, 91058 Erlangen, Germany}
\date{\today}
\begin{abstract}
The nonlinear optical properties of semiconductors near an excitonic resonance are theoretically investigated by using the macroscopic J-model [Th. \"{O}streich and A. Knorr, Phys. Rev. B {\bf 48}, 17811 (1993); {\it ibid.} {\bf 50}, 5717 (1994)] based on the microscopic semiconductor Bloch equations. These nonlinear properties cause modulational instability of long light pulses with large gain, and give rise to a self-induced transmission of short light pulses in a semiconductor. By an example of the latter well studied effect the validity of the used macroscopic model is demonstrated and a good agreement is found with both existing theoretical and experimental results.
\end{abstract}
\maketitle

\section{Introduction\label{s1}}
The generation of coherent light at frequencies which are not easily reachable by lasers, and resonant but almost lossless soliton-like light pulse propagation in a semiconductor are remarkable and useful optical phenomena. It has been recently shown~\cite{smyrnov} that the first effect can be, in particular, realized by using the modulational instability~\cite{agrawal} (MI) of a long light pulse (continuous wave) propagating in a semiconductor and resulting in the parametric growth of equidistant spectral sidebands. The second effect, the so-called self-induced transmission (SITm),~\cite{giess98} has been demonstrated both numerically and experimentally for short intense light pulses.~\cite{knorr,giess98,giess01,giess05} Both of these effects originate from the exchange Coulomb exciton-exciton interaction, which results in the nonlinear interplay of a semiconducting medium with a light pulse spectrally centered near the excitonic resonance.

Like most of the nonlinear optical phenomena in semiconductors, the mentioned effects can be theoretically investigated by using the microscopic semiconductor Bloch equations (SBEs),~\cite{lindberg,haug} which however are rather complicated for both an analytical and numerical treatment, and require large computational facilities, being based on a many-body formalism. To obtain any relatively simple analytical results either the limiting behavior of SBEs~\cite{ostr2,ostr3} or approximate but macroscopic models of these equations~\cite{ostr1,bowden} have to be considered. One of such macroscopic models, the so-called J-model, was proposed in Refs.~\onlinecite{ostr1,ostr2}, where it was demonstrated that the model sufficiently well reproduced results obtained by numerical solution of the full set of microscopic SBEs for a certain range of incident light pulse intensities. Here we apply this model to study the mentioned optical effects. Unlike previous work,~\cite{smyrnov} where we have investigated MI and solitary waves formation in semiconductors by using just a single nonlinear equation for the semiconductor polarization, the present formalism describes also the dynamics of electron-hole density (or equivalently, of the so-called inversion) in a semiconductor and thus does not have the restrictions (i.e. weak excitation condition) of the model of Ref.~\onlinecite{smyrnov}.

The structure of the present work is as follows. On the basis of a macroscopic analogue of SBEs developed within the J-model and coupled to Maxwell's equations for the electro-magnetic field of the incident pulse, we derive the system of governing equations (Sec.~\ref{s2}). Then we use this system to analyze the stability of a long light pulse with regard to small perturbations. As a result we have found a strong MI of the pulse spectrum with a very large gain. All the results were also confirmed by a direct numerical simulation of the system of governing equations without additional approximations (Sec.~\ref{s3}). To demonstrate the validity of the used model we have also applied it to describe the well studied SITm effect.~\cite{giess98} After performing the slowly varying envelope approximation (SVEA)~\cite{agrawal} both the analysis of the governing equations and their numerical simulations revealed a good agreement with known theoretical~\cite{ostr1,ostr3} and experimental~\cite{giess01,giess05} results (Sec.~\ref{s4}). CdSe was selected as a representative material in this work because it was a semiconductor utilized in known experiments on SITm.~\cite{giess98,giess01,giess05}

\section{Macroscopic semiconductor Maxwell-Bloch equations\label{s2}}
It has been demonstrated in Refs.~\onlinecite{ostr1,ostr2} that in the coherent regime, and for certain conditions, the complete system of microscopic SBEs~\cite{lindberg,haug} can be sufficiently well approximated by a set of equations for macroscopic variables that are much more suitable for an analytical treatment. Such macroscopic variables are the dimensionless complex envelope $p$ of the polarization field $\mathcal{P}(z,t)=\left\{p(z,t)\exp[ikz-i\omega t]+\cc\right\}/2$ and the dimensionless inversion $w(z,t)=2N_\mathrm{eh}(z,t)-1$ of a semiconductor, where $N_\mathrm{eh}$ is the normalized electron-hole density ($N_\mathrm{eh}\in[0,1]$).~\cite{ostr1} Here it is assumed that light is either guided or polarized, and $z$ is the longitudinal coordinate. In complex form this set of equations can be given as:
\begin{equation}\label{J_sys}
\begin{cases}
\partial_t p=-i[\Delta p + Jp(w+1)+\Omega w],\\
\partial_t w=2i[\Omega p^\ast-\Omega^\ast p],
\end{cases}
\end{equation}
where $\Delta=\omega_\rmx-\omega$ defines the detuning of the pulse central frequency $\omega$ from the 1s-exciton resonance $\omega_\rmx$, $\Omega=d_\mathrm{cv}E/\hbar$ is the Rabi frequency, while $E$ is the complex envelope of electric field $\mathcal{E}(z,t)=\left\{E(z,t)\exp[ikz-i\omega t]+\cc\right\}/2$, and $d_\mathrm{cv}$ is the interband dipole matrix element. $J$ is a phenomenological model parameter, which originates from the exchange (renormalization) terms of SBEs. In the low excitation regime, in which $\Omega\ll\omega_\rmb$ ($\omega_\rmb$ is the 1s-exciton binding frequency), $J$ was obtained on a microscopic level and was defined in Ref.~\onlinecite{ostr2}. Within the unscreened jellium model one finds $J=13\omega_\rmb/3$. We assume that the use of this definition of $J$ in system~\eqref{J_sys} is well justified, at least, for the low excitation regime.~\cite{ostr2}
This is the so-called J-model and it has been shown~\cite{ostr1} that it reproduces results obtained by numerical solution of the full system of microscopic SBEs in the range of incident pulse intensities, which corresponds to the condition $\Omega\leq\omega_\rmb$. A system similar to set~\eqref{J_sys} was also obtained from SBEs in Ref.~\onlinecite{bowden}. However, here we prefer to use the formalism of the J-model, in which, unlike Ref.~\onlinecite{bowden}, the electric field is considered separately from the exchange (renormalization) terms of SBEs, which in their turn are included in the model parameter.~\cite{ostr2}

System~\eqref{J_sys} is more general than the nonlinear equation for the macroscopic polarization field previously used in Ref.~\onlinecite{smyrnov} to investigate nonlinear optical properties of semiconductors. Set~\eqref{J_sys} can be reduced to the previous case without inversion~\cite{smyrnov} in the low excitation regime under the well-known approximation $w\simeq-1+2|p|^2$. For $J=0$ set~\eqref{J_sys} is equivalent to the optical Bloch equations (OBEs),~\cite{allen} which govern nonlinear light propagation in systems of independent two-level atoms. A more detailed discussion on a limiting behavior of system~\eqref{J_sys} solutions will be given in Sec.~\ref{s4}.

Obviously system~\eqref{J_sys} is coupled to the Maxwell equations for the electric and magnetic fields, $\mathcal{H}(z,t)=\left\{H(z,t)\exp[ikz-i\omega t]+\cc\right\}/2$. Written for envelopes without any additional approximations and in a proper dimensionless form together with set~\eqref{J_sys} they are the governing equations in this work:
\begin{equation}\label{main_sys}
\begin{cases}
\partial_x\eta=(i\omega'-\partial_T)(\psi+\lambda p)-ik'\eta,\\
\partial_x\psi=(i\omega'-\partial_T)\eta-ik'\psi,\\
\partial_T p=-i[(\Delta'-i\gamma'_\rmx)p + J'p(w+1)+\psi w],\\
\partial_T w=2i[\psi p^\ast-\psi^\ast p]-\gamma'_\rmr(w+1).
\end{cases}
\end{equation}
In the literature such systems are referred as the semiconductor Maxwell-Bloch equations (SMBEs). Here we have performed the next scalings and redefinitions: $\eta=H/H_0$, $\psi=\Omega t_0$, $x=z/z_0$, $T=t/t_0$, $\omega'=\omega t_0$, $k'=k z_0$, $\Delta'=\Delta t_0$, $J'=J t_0$, $z_0=ct_0/n$, $H_0=\hbar n/(d_\mathrm{cv}t_0)$, $\lambda=\tilde{a}t_0$, $\gamma'_\rmx=\gamma_\rmx t_0$, $\gamma'_\rmr=\gamma_\rmr t_0$, $c$ is the velocity of light in a vacuum, $n$ is the non-resonant background refractive index. $\tilde{a}=2d^2_\mathrm{cv}/(\pi a^3_0\hbar\ee\eb)$ is the photon-exciton coupling parameter, which naturally appears in Maxwell's equations after transition to the indicated dimensionless variables, $\eb=n^2$ is the bulk background dielectric constant, $\ee$ is the vacuum permittivity, and $a_0$ is the 1s-exciton Bohr radius. Parameter $\tilde{a}$ defines the width of forbidden frequencies region in the dispersion relation of a photon-exciton coupled state.~\cite{haug} The exciton damping $\gamma_\rmx$ and relaxation $\gamma_\rmr$ parameters are introduced here phenomenologically, and $t_0$ is an arbitrary time scaling parameter, which will be properly chosen in  the following sections.

\section{Modulational instabilities\label{s3}}
In analogy with Ref.~\onlinecite{smyrnov} using SMBEs~\eqref{main_sys} we perform the MI analysis~\cite{agrawal} - we analyze the linear stability of a long light pulse propagating in a semiconductor with respect to small perturbations. For this purpose we substitute the perturbed fields and polarization envelopes $\{\psi;\eta;p\}(x,T)=(\{\psi_0;\eta_0;p_0\}+\{a;g;b\}(x,T))\exp[iqx]$ as well as the perturbed inversion $w(x,T)=w_0+d(x,T)$ into system~\eqref{main_sys}. After that we define the small perturbations as $\{a;g;b\}(x,T)=\{a_1;g_1;b_1\}\exp[i\kappa x - i\delta T]+\{a_2;g_2;b_2\}\exp[i\delta T-i\kappa^\ast x]$, while $d(x,T)=2d_0\re\exp[i\kappa x-i\delta T]$ because the inversion is a real function. Then, requiring that system~\eqref{main_sys} is solvable, we obtain within the first-order perturbation theory~\cite{agrawal} the dispersion relation $\kappa(\delta)$ for perturbations:
\begin{widetext}
\begin{equation}\label{mi_matrix}
\left|
  \begin{array}{ccccc}
    (\omega'+\delta)^2-(q+\kappa)^2 & 0 & \lambda(\omega'+\delta)^2 & 0 & 0\\
    0 & (\omega'-\delta)^2-(q^\ast-\kappa)^2 & 0 & \lambda(\omega'-\delta)^2 & 0 \\
    p^\ast_0 & -p_0 & -\psi^\ast_0 & \psi_0 & \delta/2\\
    w_0 & 0 & \Delta'+J'(w_0+1)-\delta & 0 & J'p_0+\psi_0 \\
    0 & w_0 & 0 & \Delta'^\ast+J'(w_0+1)+\delta & J'p^\ast_0+\psi^\ast_0 \\
  \end{array}
\right|=0,
\end{equation}
\end{widetext}
where $q=\omega'[1-\lambda w_0/(\Delta'+J'(w_0+1))]^{1/2}$, $\kappa$ is the perturbation wavenumber, and $\delta$ is the relative perturbation frequency. By using the relation between the steady-state amplitudes $p_0=-\psi_0w_0/(\Delta'+J'(w_0+1))$ and a conservation law $w^2+4|p|^2=1$, which follows from~\eqref{J_sys}, one can obtain a $4$-th order algebraic equation for $w_0$:
\begin{equation}\label{ww}
(w^2_0-1)|\Delta'+J'(w_0+1)|^2+4w^2_0|\psi_0|^2=0.
\end{equation}
This defines $w_0$ for given $\psi_0$ and system parameters. Among its solutions we select the real ones (either two or four): $w_{0}\in(-1;1)$. In the most of the cases there are only two real solutions - a negative and a positive one, and they correspond to a semiconductor, which was relatively weakly or strongly excited before the incident pulse arrival. However, of course, for certain combinations of parameters there exist four different real solutions, each of which corresponds to a different excitation regime. Therefore, setting certain incident pulse intensity and detuning, and choosing one of the mentioned excitation regimes, we completely determine the dispersion relation~\eqref{mi_matrix} as all the other parameters are defined by a medium. It is important to note, that, like in Ref.~\onlinecite{smyrnov}, the excitonic damping is naturally taken into account in Eq.~\eqref{mi_matrix} by a proper final redefinition $\Delta'\rightarrow\Delta'-i\gamma'_\rmx$. After this the steady-state amplitudes and parameter $q$ also become complex, resulting in the appearance of conjugated terms in Eq.~\eqref{mi_matrix} after the MI analysis.
\begin{figure}[tb]
\includegraphics[width=8.4cm]{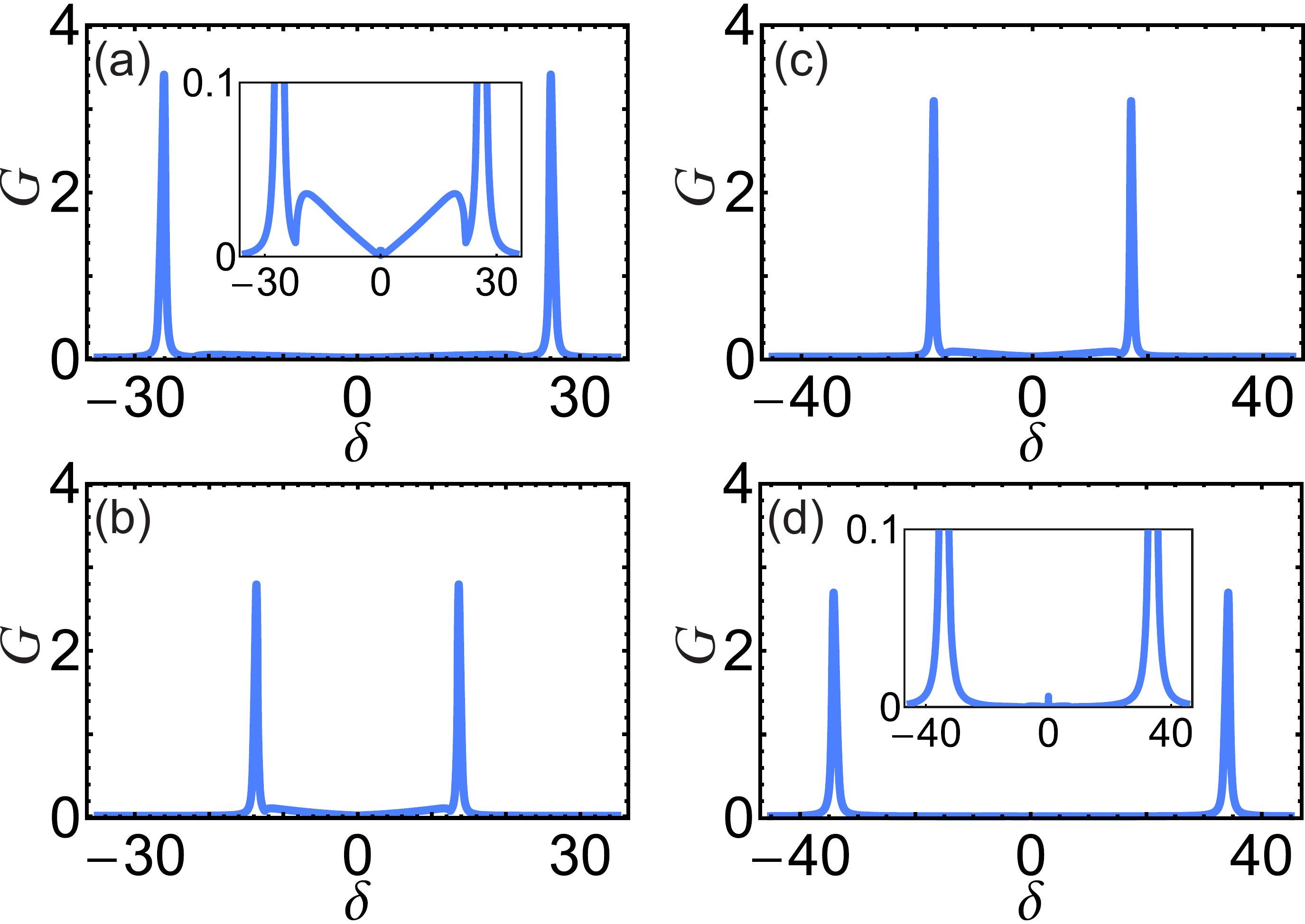}
\caption{The MI gain $G(\delta)$ spectrum according to Eq.~\eqref{mi_matrix} for $\psi_0=5$, $\Delta'=1.5$, and $\gamma'_\rmx=0.267$. (a) For $J'=65.4$ and $w_0\gtrsim-1$. (b) The same $w_0$ as in (a) but smaller $J'=6.54$. (c) For $J'=16.35$ and $w_0\gtrsim-1$. (d) The same $J'$ as in (c) but different $w_0\lesssim1$. Insets at (a) and (d) show the same gain spectra but for a smaller vertical scale. $\omega'$ is positioned at the origin.}\label{fig_mi}
\end{figure}

In the frequency regions, for which $\im{\kappa(\delta)}<0$ is valid, the exponential growth of perturbations takes place, defining the MI gain spectrum $G(\delta)=2\max\left|\im{\kappa(\delta)}\right|$. Since Eq.~\eqref{mi_matrix} is of $4$-th order in $\kappa$, among its solutions we select, for a given $\delta$, the one with the maximum absolute value of the gain. The MI gain spectra for different $J'$ are given at Fig.~\ref{fig_mi}, where high resonant gain peaks clearly manifest themselves. The nature of these resonant peaks, which are strongly affected by the precise value of $\gamma'_\rmx$, is discussed in Ref.~\onlinecite{smyrnov}. A possibility to vary $J'$ due to screening effects is shown in Sec.~\ref{s4}, while here we only demonstrate how the change of $J'$ can influence the MI gain spectrum (compare Figs.~\ref{fig_mi}(a) and~\ref{fig_mi}(b)). Increase of $J'$ can lead to a shift of MI gain peaks farther from the incident pulse central frequency $\omega'$ and to their slight gain enhancement (Fig.~\ref{fig_mi}(a)). At Figs.~\ref{fig_mi}(c) and~\ref{fig_mi}(d) for a fixed $J'$ (and other parameters) we compare the MI gain spectra in different excitation regimes. In the regime of pre-excited semiconductor ($w_0\lesssim1$, Fig.~\ref{fig_mi}(d)) the MI gain peaks are shifted essentially farther from $\omega'$ than those in the case of initially unexcited medium ($w_0\gtrsim-1$, Fig.~\ref{fig_mi}(c)). However, depending on the selected parameters the opposite behavior is also possible. It is also worth mentioning that in the case when all the four solutions of Eq.~\eqref{ww} are real, one of them usually corresponds to a regime with zero gain - the pulse is stable with respect to small perturbations.

\begin{figure}[tb]
\includegraphics[width=6.3cm]{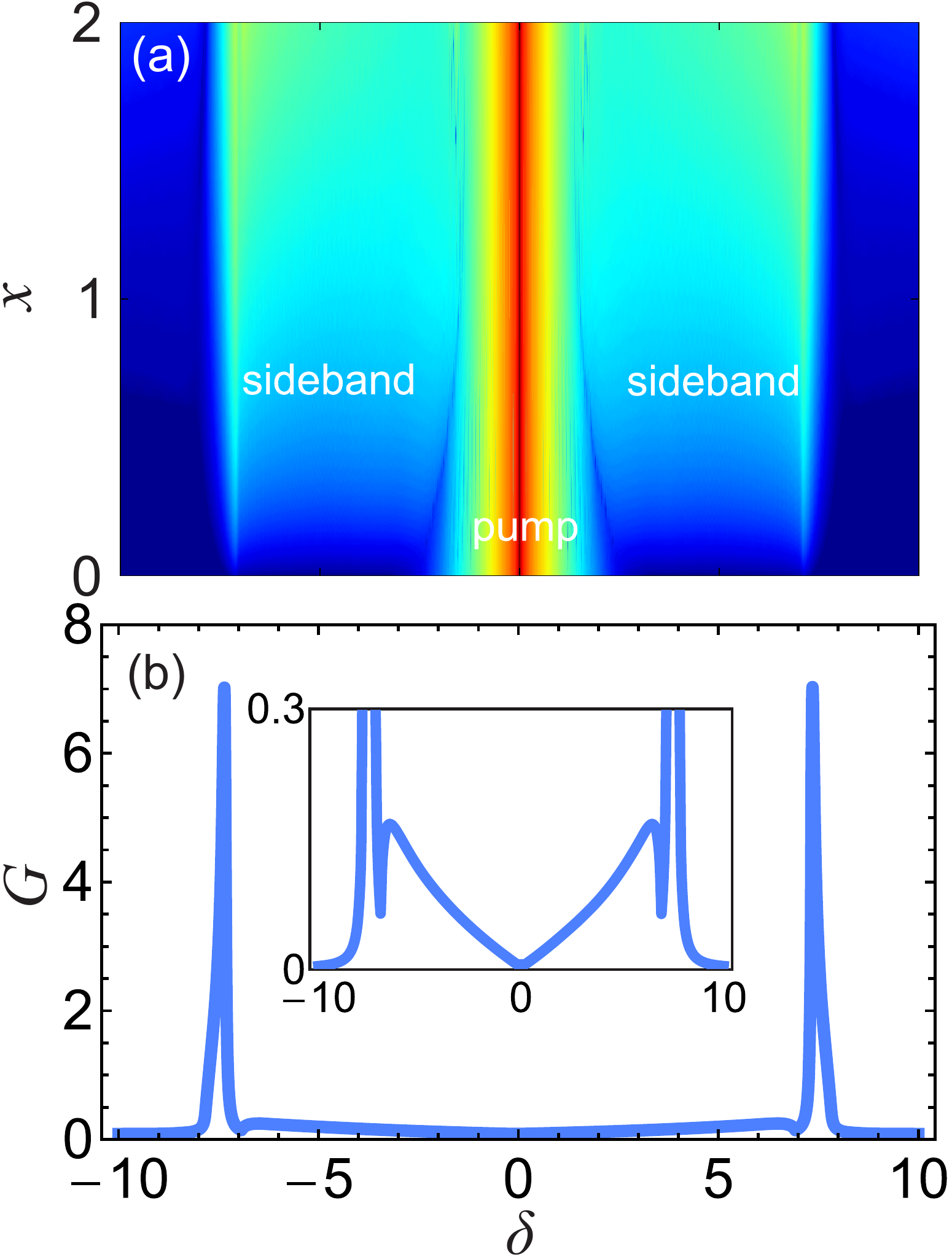}
\caption{(a) Numerically simulated long pulse spectral evolution according to system~\eqref{main_sys}, the initial pulse is $\psi(x=0,T)=\psi_0\exp\left[-(2T/T_\mathrm{w})^{2m}\right]$, $\psi_0=3$, $T_\mathrm{w}=200$, $m=20$, while the other parameters are: $\Delta'=1.5$, $J'=1.96$, and $\gamma'_\rmx=0.05$. (b) The corresponding analytical prediction on the MI gain maxima positions by Eq.~\eqref{mi_matrix}. Inset at (b) shows the same gain spectrum but for a smaller scale of $G$. $\omega'$ is positioned at the origin.}\label{fig_misim}
\end{figure}
To confirm results of the mentioned analysis we have also performed direct numerical simulations of the full system~\eqref{main_sys} by using a $4$-th order Runge-Kutta algorithm with the following initial conditions: $\psi(x=0,T)$ is a wide super-Gaussian pulse, $\eta(x=0, T)=\psi(x=0, T)$, $p(x=0,T=0)=0$, and $w(x=0,T=0)=-1$, what means that the medium is not excited before the field arrival (see Fig.~\ref{fig_misim}(a)). A comparison with the respective analytical prediction on the MI gain maxima positions indicates an excellent agreement (Fig.~\ref{fig_misim}(b)). The regime of a preliminarily excited medium was also realized numerically by using the initial condition $w(x=0,T=0)\lesssim1$ and demonstrated qualitatively similar spectral evolution in accordance with the analytical prediction on the sidebands positions. At this point for calculations we should specify the representative material, which in this work is CdSe. For this semiconductor at low temperatures (e.g., 8~K in Ref.~\onlinecite{giess01}, 2~K in Ref.~\onlinecite{pantke}), at which excitonic effects are essential:~\cite{giess01,pantke} $\hbar\omega_\rmx=1835$~meV, $\hbar\omega_\rmb=15$~meV, $\hbar\tilde{a}=1$~meV, $\hbar\gamma_\rmx=0.265$~meV, $\gamma_\rmr<\gamma_\rmx$ according to Ref.~\onlinecite{pantke}, $a_0\simeq5.32$~nm, $d_\mathrm{cv}/e\simeq0.343$~nm, and $\eb=9$. Now we choose $t_0=1/\tilde{a}\simeq0.67$~ps and for simulations remove a large number $\omega'\sim\omega_\rmx/\tilde{a}$ from system~\eqref{main_sys} by multiplying its first two equations by $\lambda/\omega'$ and performing a space rescaling $z_0\rightarrow z_0\lambda/\omega'\sim c/(n\omega)\simeq0.04~\mathrm{\mu m}$. In accordance with this scaling one can estimate, like in Ref.~\onlinecite{smyrnov}, from Fig.~\ref{fig_mi} the semiconductor MI gain $\sim100~\mu\mathrm{m}^{-1}$, which is huge in comparison with that of, for example, silica optical fibers ($\sim10~\mathrm{km}^{-1}$).~\cite{agrawal}

It is interesting to note, that a MI gain spectrum qualitatively similar to the considered ones can be obtained even for $J=0$ (the case of OBEs),~\cite{allen} something that was impossible in the model of Ref.~\onlinecite{smyrnov}.

\section{Self-induced transmission solitons\label{s4}}
To demonstrate the validity of the used macroscopic model we now apply it to a well studied nonlinear optical effect. The SITm effect in semiconductors~\cite{giess98} is an analogue of the known self-induced transparency (SIT) effect in idealized systems of two-level atoms.~\cite{allen,call} Although both of them consist in almost lossless soliton-like propagation of short (subpicosecond in semiconductors) light pulses in a resonant medium, they essentially differ, what is discussed in details, for example, in Refs.~\onlinecite{knorr,giess98,giess01,giess05}. Here we only briefly remind that due to strong influence of Coulomb exchange interactions between the Wannier excitons in semiconductors, which are not inherent to electrons in atomic systems, the Rabi frequency of the incident field approximately doubles~\cite{binder} in a semiconducting medium, when it is about $\Omega\sim\omega_\rmb$.~\cite{ostr1,ostr3} In particular, this results in a substantial deviation from the so-called area theorem, which is known from the theory of a light pulse propagation in systems of two-level atoms and mainly states that only for incident pulse of an area $\theta=2\int^{\infty}_{-\infty}|\psi|dT|_{x=0}$ equal to $2\pi$ SIT effect takes place - pulse is stable during the propagation. A pulse with an area, which is integer multiple of $2\pi$, should undergo a breakup into $2\pi$-pulses.~\cite{allen} Such phenomena have been also found in semiconductors but for pulses with area $\theta\simeq\pi$ and their respective multiples.~\cite{koch,giess01,giess05} Such experimental and numerical observations (based on microscopic SBEs) have never been explained in clear physical terms in the literature - from which one of the motivations of the present work originates.

\begin{figure}[tb]
\includegraphics[width=8.4cm]{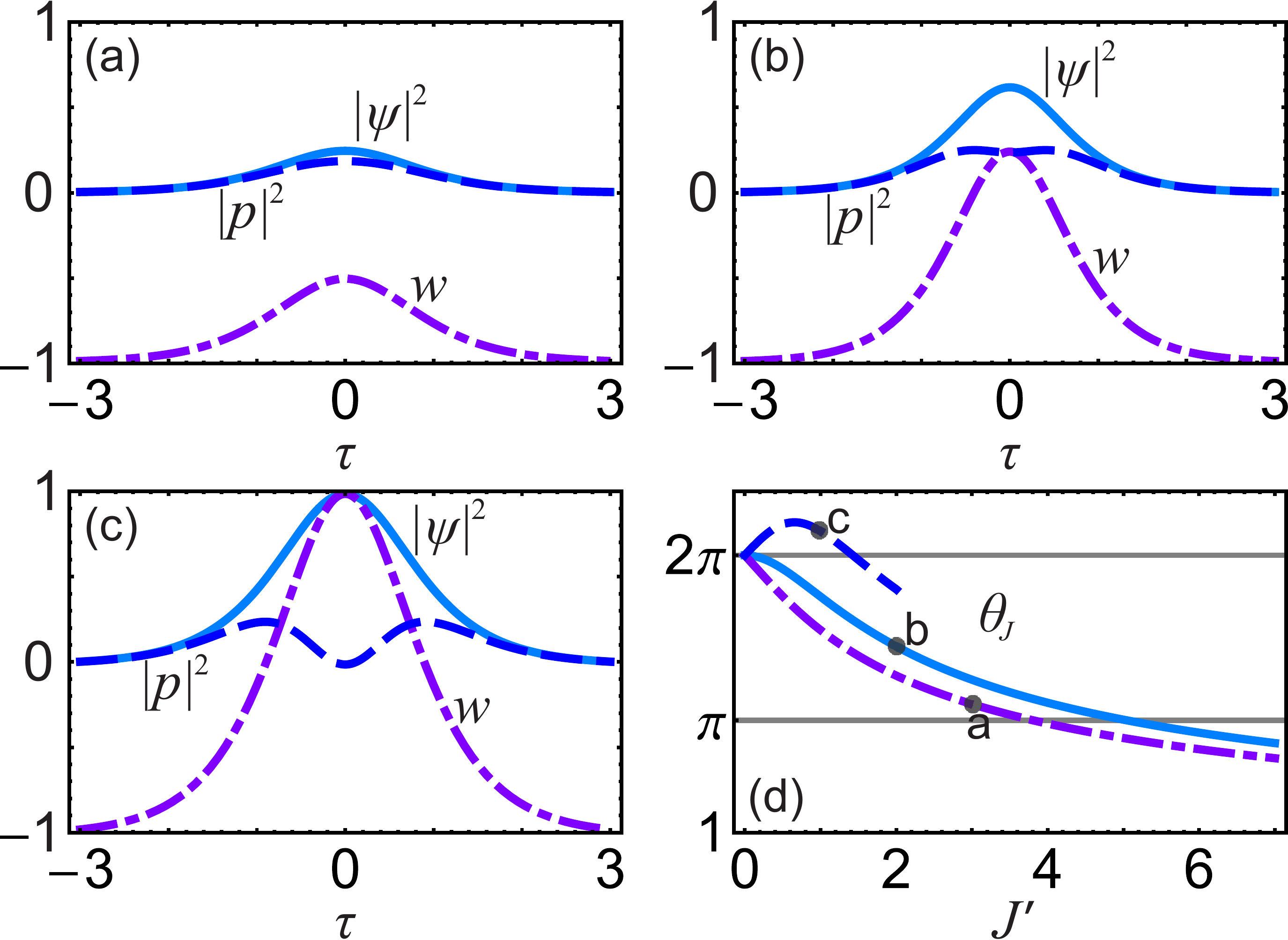}
\caption{(a) Temporal shapes of $|\psi|^2$ (solid), $|p|^2$ (dashed), and $w$ (dot-dashed line) in accordance with Eq.~\eqref{sit_sol} for $\Delta'=1$ and $J'=3$. (b) The same functions but for $\Delta'=0$ and $J'=2$. (c) The same but for $\Delta'=-1$ and $J'=1$. (d) Dependence of the pulse area $\theta_J$ on the value of $J'$ for $\Delta'=\{-1;~0;~1\}$ - dashed, solid, and dot-dashed lines, respectively. Grey dots indicate the pulse areas for the considered cases. The parameter space restrictions of soliton~\eqref{sit_sol} are clearly demonstrated by the example of $\Delta'=-1$.}\label{fig_psi}
\end{figure}

Here we demonstrate the latter effect using system~\eqref{main_sys}, changing the variable $\tau=T-x/V$ ($V$ is the dimensionless pulse group velocity), and applying the SVEA.~\cite{agrawal} SMBEs~\eqref{main_sys} transform in this way into system:
\begin{equation}\label{svea_sys}
\begin{cases}
\partial_\tau\psi=-ip,\\
\partial_\tau p=-i[(\Delta'-i\gamma'_\rmx)p + J'p(w+1)+\psi w],\\
\partial_\tau w=2i[\psi p^\ast-\psi^\ast p]-\gamma'_\rmr(w+1),
\end{cases}
\end{equation}
where we set $t_0=\sqrt{\alpha}$, and $\alpha=2(\tilde{a}\omega)^{-1}(1/V-1)>0$ as $V<<1$.~\cite{giess01}
Following Ref.~\onlinecite{afan}, where light propagation in a dense medium of interacting dipoles has been investigated by using a system analogous to set~\eqref{svea_sys}, we briefly reproduce an exact solution of Eqs.~\eqref{svea_sys} in the coherent limit ($\gamma_\rmx,\gamma_\rmr=0$). Evidently, when using the first equation in set~\eqref{svea_sys} the last one becomes integrable, and hence $w=-1+2|\psi|^2$. Therefore, the whole set~\eqref{svea_sys} reduces to a single equation for $\psi$ only:
\[
\partial^2_\tau\psi+i(\Delta'+2J'|\psi|^2)\partial_\tau\psi+(2|\psi|^2-1)\psi=0.
\]
Looking for its solution in the form $\psi(\tau)=\widetilde{\psi}(\tau)\exp[i\varphi(\tau)]$, $\widetilde{\psi}\equiv|\psi|$, we finally come to a couple of equations:
\begin{subequations}
\begin{equation}\label{cqe1}
\partial^2_\tau\widetilde{\psi}+A\widetilde{\psi}^5+B\widetilde{\psi}^3-C\widetilde{\psi}=0,
\end{equation}
\begin{equation}\label{cqe2}
\varphi=-\frac{\Delta'\tau}{2}-\frac{J'}{2}\int^{\tau}_{-\infty}\widetilde{\psi}^2d\tau,
\end{equation}
\end{subequations}
where $A\equiv3J'^2/4$, $B\equiv\Delta'J'+2$, and $C\equiv1-\Delta'^2/4$. Eq.~\eqref{cqe1} contains the so-called cubic-quintic nonlinearity, it is integrable and has a solitonic solution:~\cite{kivshar}
\begin{equation}\label{sit_sol}
\widetilde{\psi}=\frac{2\sqrt{C/B}}{\left(1+\sqrt{1+16AC/(3B^2)}\cosh[2\sqrt{C}\tau]\right)^{1/2}}.
\end{equation}
The parameter space for this soliton is $|\Delta'|<2$, $\Delta'>-2/J'$ - it can form only if the incident pulse is spectrally centered very close to the excitonic resonance. The area of soliton~\eqref{sit_sol} $\theta_J=2\int^{\infty}_{-\infty}\widetilde{\psi}~d\tau$ tends to $2\pi$ only if $J'\rightarrow0$ and $\widetilde{\psi}\rightarrow\sqrt{C}\sech[\sqrt{C}\tau]$, otherwise almost for the whole range of parameters $\theta_J<2\pi$ (see Fig.~\ref{fig_psi}). This, in accordance with known theories~\cite{koch,knorr} and experiments,~\cite{giess01,giess05} clearly demonstrates within the used model that exchange Coulomb interactions between excitons, from which parameter $J$ actually originates,~\cite{ostr2} can reduce the area of a SITm soliton in a semiconductor (see Fig.~\ref{fig_psi}(d), an explanation on why $J$ can vary is given further in this section).

\begin{figure}[tb]
\includegraphics[width=8.4cm]{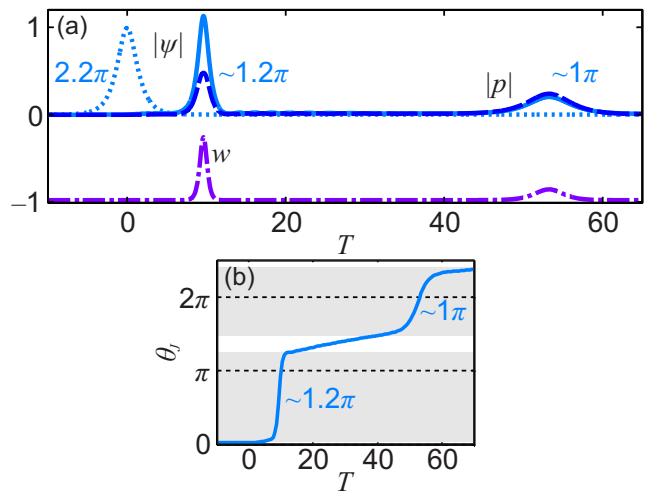}
\caption{(a) Numerical demonstration of a pulse breakup under propagation by using system~\eqref{svea_sys}. The electric field $|\psi|$ (solid), the polarization $|p|$ (dashed) and the inversion $w$ (dot-dashed line) after a propagation of $x=60$ are shown. The initial double pulse (dotted line) is $\psi(x=0,T)=2\widetilde{\psi}(x=0,T)\exp[i\varphi(x=0,T)]$ in accordance with Eqs.~\eqref{cqe2} and~\eqref{sit_sol}, while the parameters are $\Delta'=1$ and $J'=3$ (similar to those at Fig.~\ref{fig_psi}(a)). The areas $\theta_J$ of electric field pulses are indicated. (b) Changes in the total $|\psi|$-pulse area $\theta_J$ with time after the propagation. The regions, which correspond to the areas of formed pulses, are indicated by the grey bars.}\label{fig_sit}
\end{figure}

To support the above mentioned analytical conclusions we have also performed direct numerical simulations of system~\eqref{svea_sys} after returning to the $x,T$-variables, and by employing a numerical algorithm similar to that described in Sec.~\ref{s3}. In Fig.~\ref{fig_sit}(a) a propagating pulse of a doubled area of $\theta_J\simeq2.2\pi$ [parameters are given at Fig.~\ref{fig_psi}(a)] undergoes a breakup in two well separated pulses of areas $\theta_J\simeq1.2\pi$ and $\simeq1\pi$ [see Fig.~\ref{fig_sit}(b)], what is in good accordance with the prediction on a SITm soliton shape and area $\sim1.1\pi$ given by Eq.~\eqref{sit_sol} (compare with Figs.~\ref{fig_psi}(a) and~\ref{fig_psi}(d)). We should also notice that the soliton shown at Fig.~\ref{fig_psi}(c) ($\Delta'=-1$, $J'=1$) was unstable under the propagation.

Now, in order to perform a more detailed limiting analysis of solution~\eqref{sit_sol} and to obtain some quantitative estimates we return to the microscopic definition of model parameter $J$, which, as we have already mentioned, is given in Ref.~\onlinecite{ostr2} ($J=\beta_1/2$, formula 12) for the low excitation regime $\Omega\ll\omega_\rmb$. In regimes of stronger excitation, for higher electron-hole densities the screening of interparticle Coulomb interactions in a semiconductor becomes important.~\cite{lindberg} Here we assume that the microscopic definition of $J$ from Ref.~\onlinecite{ostr2} remains valid even in those regimes, but for calculation of $J$ within the jellium model instead of the bare Coulomb interaction potential and the respective 1s-exciton hydrogenic wave function we use a proper screened potential and the corresponding ground-state eigenfunction of the Schr\"{o}dinger equation. We utilize a natural, statically screened Coulomb interaction potential - the Yukawa potential:~\cite{haug} $v(r)=e^2(4\pi\ee\eb)^{-1}\exp[-\kappa_\rms r]/r$, where $r$ is the radius-vector modulus, and $\kappa_\rms$ is the screening wave number. Accurate analytic approximations of the respective ground-state eigenvalue and eigenfunction are given, for example, in Ref.~\onlinecite{liverts}. Phenomena related with excitons take place at rather low temperatures, therefore all the experiments on the SITm in semiconductors were performed in this range of temperatures (e.g., 8~K, like in Refs.~\onlinecite{giess01,giess05}). That is why we use here the Thomas-Fermi form of the screening wave number~\cite{haug} $\kappa_\rms=[\mu e^2(3N_\rmx/\pi)^{1/3}/(\pi\ee\eb\hbar^2)]^{1/2}$, where $\mu$ is the exciton reduced effective mass (it is $0.0969~m_\mathrm{e}$ for CdSe),~\cite{giess01} and $N_\rmx$ is the total density of excitons (e-h pairs) generated by an incident pulse. We estimate it here as $N_\rmx\sim I/(cn\hbar\omega_\rmx)$, $I=cn\ee(\hbar\Omega)^2/(2d^2_\mathrm{cv})$ is the incident pulse intensity.~\cite{agrawal} In this way we come to an approximate but natural and explicit relation between the model parameter $J$ and the incident field Rabi frequency - $J$ can essentially decrease, when $\Omega$, $N_\rmx$, and $\kappa_\rms$ are large enough and the screening effects become significant.

The limiting behavior of the SITm solitons is thereby straightforward. For the high excitation regime ($\Omega\gg\omega_\rmb$) the screening is strong, electron-hole pairs are almost decoupled, and $J\rightarrow0$, while $\theta_J\rightarrow2\pi$ in complete accordance with the results of analysis of microscopic SBEs limiting behavior in Ref.~\onlinecite{ostr1}. For the low excitation regime ($\Omega\ll\omega_\rmb$) the screening effects are negligible and $J\rightarrow13\omega_\rmb/3$, while $\theta_J$ becomes even smaller than unity, again in accordance with the results of analysis of the respective limiting behavior of SBEs.~\cite{ostr3} However, we should note that in the used model $J$ remains finite and $\theta_J\lesssim1$ even for $\Omega$ below the low excitation regime, while in Ref.~\onlinecite{ostr3} in this case $\theta_J\ll1$. This indicates a restriction of the used model - $\Omega$ should not be vanishing. Here and further in this section calculations are performed by using data on $t_0=\sqrt{\alpha}=(\tau^{-2}_\mathrm{w}+\Delta^2/4)^{-1/2}=0.8$~ps ($\tau_\mathrm{w}$ is the soliton temporal width) and $\Delta=0$, taken from the experiment on the SITm in CdSe.~\cite{giess01}

In the intermediate case $\Omega\sim\omega_\rmb$ (for CdSe this corresponds to $I\sim750~\mathrm{MW/cm^2}$), for which an approximate doubling of the incident field Rabi frequency should occur,~\cite{ostr1,ostr3} we estimate $J$ and $\theta_J$ using typical incident pulse intensities employed in experiments on the SITm in CdSe:~\cite{giess01,giess05} $I\sim100~\mathrm{MW/cm^2}$. According to the developed formalism this yields $N_\rmx\simeq3.8\cdot10^{15}\mathrm{cm^{-3}}$, $1/\kappa_\rms\simeq8.94$~nm, $J\simeq0.236~\omega_\rmb$, and hence $\theta_J\simeq1.08\pi$ (e.g., $\theta_J\in[0.82,1.48]\pi$ for $I\in[50,200]~\mathrm{MW/cm^2}$). Therefore, in good accordance with the known experiments~\cite{giess01,giess05} for the considered intensities range the SITm effect indeed takes place for incident pulses of areas $\theta_J\simeq\pi$.

\section{Conclusions\label{s5}}
In summary, a strong resonant MI effect in semiconductors has been demonstrated analytically on the basis of a known simple semiclassical macroscopic model,~\cite{ostr1,ostr2} which originates from microscopic SBEs. New ways to shift the MI gain peaks, and therefore the generated frequencies have been discussed. The analytical results are completely confirmed by direct numerical simulations of the governing equations. The validity of the used macroscopic model has been shown by studying the SITm effect in semiconductors, for which a good agreement with both existing theoretical~\cite{ostr1,ostr3} and experimental~\cite{giess01,giess05} results has been found. Further investigation of nonlinear light-semiconductor interplay beyond the SBEs formalism (e.g., see Ref.~\onlinecite{schafer}) can lead to a revelation of novel aspects of the studied phenomena.

\begin{acknowledgments}
This work was supported by the German Max Planck Society for the Advancement of Science (MPG).
\end{acknowledgments}

\end{document}